\documentclass[12pt]{article}
\usepackage{graphicx}
\usepackage{subfigure}
\usepackage[dvips]{color}
\newcommand{\beq}{\begin{equation}}
\newcommand{\eeq}{\end{equation}}
\newcommand{\bea}{\begin{eqnarray}}
\newcommand{\eea}{\end{eqnarray}}

\newcommand{\overlrarrow}[1]{\vbox{\ialign{##\cr\cr
                  \leftrightarrowfill\crcr\noalign{\kern-1pt\nointerlineskip}
                  $\hfil\displaystyle{#1}\hfil$\crcr}}}

\begin{document}
\begin{titlepage}
\begin{flushleft}
       \hfill                     
       \hfill                       FIT HE - 16-01 \\
\end{flushleft}
\vspace*{3mm}
\begin{center}
{\bf\LARGE Holographic Schwinger Effect and \\
\vspace*{3mm}
 Chiral condensate in SYM Theory}

\vspace*{5mm}
\vspace*{2mm}
\vspace*{5mm}
{\large Kazuo Ghoroku${}^{\dagger}$\footnote[1]{\tt gouroku@dontaku.fit.ac.jp},
Masafumi Ishihara${}^{\ddagger}$\footnote[2]{\tt masafumi@wpi-aimr.tohoku.ac.jp},
%
}\\

{${}^{\dagger}$Fukuoka Institute of Technology, Wajiro, 
Higashi-ku} \\
{
Fukuoka 811-0295, Japan\\}
{
${}^{\ddagger}$ WPI-Advanced Institute for Materials Research (WPI-AIMR),}\\
{
Tohoku University, Sendai  980-8577, Japan\\}

\end{center}

\begin{abstract}

We study the instability, for the supersymmetric Yang-Mills (SYM) theories, caused by the external electric field 
through the imaginary
part of the action of the D7 probe brane, which is embedded in the background of type IIB theory. 
This instability is related to the Schwinger effect,
namely to the quark pair production due to the external electric field,
for the $SU(N_c)$ SYM theories. 
In this holographic approach, 
it is possible to calculate the Schwinger effect for various phases of the theories.
Here we give the calculation for ${\cal N}=2$ SYM theory and the analysis is extended to
the finite temperature deconfinement and  the
zero temperature confinement phases of the Yang-Mills (YM) theory. 
By comparing the obtained production rates with the one of the supersymmetric case, 
the dynamical quark mass is estimated and we find how it varies with 
the chiral condensate. Based on this analysis, we give a speculation on the extension of the 
Nambu-Jona-Lasinio model to the finite temperature YM theory, and four fermi coupling
is evaluated in the confinement theory.

\end{abstract}
\end{titlepage}

\section{Introduction}


The phenomenon related to flavor quarks, which is identified with the ${\cal N}=2$ hypermultiplet, in suersymmetric Yang-Mills(SYM) theory  
has been studied by embedding D7 brane(s) as a probe
in the D3 stacked background of the type IIB theory \cite{KK}-\cite{CNP}. 
The profile of the D7 brane 
provides us information about chiral condensate 
of a quark in the dual ${\cal N}=2$ SYM theory. 

In this direction, the "electro-magnetic" properties of the system
have been studied by imposing
the external electro-magnetic field of $U(1)_B$ on the system, where the charge of the current
corresponds to the baryon number (see refs.\cite{KO, Oba}). 
Many properties of the system have been cleared in this setting through the D7 branes.
Recently, in the SYM theory, the Schwinger effect has been studied according to 
the idea that the D7 brane 
action $S_{D7}$ can be related to the Euler-Heisenberg Lagrangian ${\cal L}(E)$ as follows \cite{Hashi},
\beq\label{D7}
    S_{D7}=-\int d^4x {L}(E)\, ,
\eeq
where the internal space of the D7 world volume is integrated out and 
$E$ denotes the external electric 
field imposed on the system. 
In this context, the Schwinger effect in SYM theory has been given by  
the imaginary part of this Lagrangian, Im${L}(E)$ \cite{Hashi} - \cite{Wu}.

On the other hand, as shown in \cite{KO}, 
this imaginary part can be removed by introducing
an appropriate electric current which brings the system to a non equilibrium steady state.
This fact implies that Im${L}(E)$, which is given for a state without the electric current, 
is regarded as the transition probability from a false vacuum 
without any current 
to the non equilibrium steady state with an appropriate current.
So it may be reasonable to regard Im${L}(E)$ as the pair production rate of  the positive and negative charges,
namely the quark and anti-quark. In this sense, this process can
be considered as the Schwinger effect \cite{Sch}.

\vspace{.3cm}
The pair production rate from (\ref{D7}) is obtained as follows. 
First, embed D7 probe brane in a
given bulk which corresponds to the vacuum state of the dual theory.
Then evaluate Im${ L}$ from (\ref{D7}) after imposing an external electric field $E$. 
Im${L}(E)$ is obtained for $E\ge E_c$ where $E_c$ is determined by the theory.  

Here we remember the Schwinger's one-loop formula which is given in 
the four dimensional (4D) quantum electrodynamics (QED).
In the case of the D7 brane embedded in the AdS$_5\times S^5$ bulk, 
a hypermultiplet (one fermion and two scalars) is considered in the $SU(N_c)$ ${\cal N}=2$
supersymmetric dual gauge theory. Thus for the hypermultiplet with the mass $m$, 
the Schwinger's formula would be considered by supposing the tunneling process
as in the QED. 
The formula 
for Im${ L}$ obtained according to the method mentioned above, is not however 
equivalent to the Schwinger's formula
due to the reason that the production rates are obtained via vacuum decay,
which is not a tunneling.

We show that
this lower bound $E_{c}$ mentioned above is needed to remove the attractive
quark potential, which increases linearly in the short range distance between 
the pair produced quark and the anti-quark in the SYM theory  
 \cite{KMMW}. 
Then enough repulsive force coming from the external field is
necessary to overcome this attractive force and to separate the pair produced quarks.
Thus, we could find a stable electric current under an enough strong $E$. 
The 
value of $E_c$ reflect the dynamics of the dual theory.

It is interesting to see the dynamical properties of the various $SU(N_c)$ gauge theories
through the Schwinger effect given by (\ref{D7}). Our purpose is to investigate this point. 
The calculations are extended to the YM theories in the
finite temperature deconfinement phase and in a zero temperature confinement phase. 

\vspace{.2cm}
The dynamical properties of these two theories are complicated and dependent on the parameters
of the theories. Here we study how the production rate $\Gamma$ depends on the chiral condensate 
$\langle\bar{\Psi}\Psi\rangle$ and its relation with the effective quark mass $m_q^{eff}$.
As for the $m_q^{eff}$, we estimate it by comparing the production rate obtained for
the supersymmetric theory, which is dual to AdS$_5\times S^5$,
with the one for the above two non-supersymmetric theories. The mass shift
in the non-supersymmetric theories is measured through the pair production rate.
Then, a speculation on the NJL model of QCD is given, and an extension of the NJL to finite temperature
theories is proposed.

\vspace{.3cm}
In \cite{SZ}, the pair production rate of the W bosons has been studied as a holographic
Schwinger effect 
in terms of a probe D3 brane and a  
string in the D3 background.
In this case, the production rate is obtained in terms of the tunneling process.
An interesting point, in this calculation, is the
existence of the second critical electric field
which is needed for the tunneling in a theory of confining phase
\cite{SY3}.  
 In the present case of D7/D3 system, this kind of production rate would be
obtained from the real part of D7 action given above. We will discuss on this point very briefly
in the article.

\vspace{.2cm}
In the next section, we give a brief review of our D3/D7 brane model for a 
finite temperature Yang-Mills (YM) theory. In the section 3, how to calculate the production rate
of the quark pair by using (\ref{D7})
as a Schwinger effect is shown. In the section 4, the production rate is calculated
for the YM  theory whose vacuum state is in the finite temperature deconfinement phase.
At first, the production rate is given for the case of the massive quark, and then the effect of the 
temperature and chiral condensate are examined. The effective quark mass is estimated by relating 
the production rate for the massive quark to the one obtained for the supersymmetric theory which has  zero chiral condensate. Then
the results are discussed by supposing the relation to the NJL model of QCD.
In the next section, the parallel analysis is performed for the theory whose vacuum
state is in the confinement
and broken chiral symmetry phase. We find a good relation between the NJL model and our 
resultant formula for the effective quark mass. Summary and discussions
are given in the final section.



\section{D3/D7 model and D7 Embedding}

We study the Schwinger effect for the ${\cal N}=4$ SYM theory coupled to ${\cal N}=2$ hypermultiplet, whose holographic dual
is given by  D3/D7 branes system in the type IIB string theory. 

Here, as a prototype of the model, we show
the case of the finite temperature deconfinement phase. Its background metric is given by
the AdS$_5$ Schwarzschild  $\times S^5$ ,  which is written as
\begin{equation}
ds^2=\frac{r^2}{R^2}\left(-f^2(r)dt^2+(dx^i)^2\right)+\frac{1}{f^2(r)}\frac{R^2}{r^2}dr^2+R^2d\Omega_5^2, \label{adsbh}
\end{equation}
where $R^4=4\pi {\alpha'}^2 N_c$ and 
\begin{equation}
f(r)=\sqrt{1-\left(\frac{r_T}{r}\right)^4}\, ,
\end{equation}
and the temperature $T$ is given by $T=\frac{r_T}{\pi R^2}$.
The embedding of the D7 brane is performed according to \cite{GSUY} by
rewriting the six dimensional part of the above metric (\ref{adsbh}) as
\begin{equation}
\frac{1}{f^2(r)}\frac{R^2}{r^2}dr^2+R^2d\Omega_5^2=\frac{R^2}{U^2}\left(d\rho^2+\rho^2d\Omega_3^2+(dX^8)^2+(dX^9)^2\right)
\end{equation}
where
\begin{equation}
U(r)=r\sqrt{\frac{1+f(r)}{2}}\, ,
\end{equation} 
and 
\begin{equation}
U^2=\rho^2+(X^8)^2+(X^9)^2.
\end{equation}
Then the DBI action of a D7-brane is given as
\begin{equation}
S_{D7}=-\tau_7\int d^8\xi \sqrt{-\det\left(g_{ab}+2\pi\alpha'F_{ab}\right)}\, , \label{sd7dbi}
\end{equation}
with the following induced metric $g_{ab}$,
\begin{equation}
ds_8^2=g_{ab}\xi^a\xi^b=\frac{r^2}{R^2}\left(-f^2(r)dt^2+(dx^i)^2\right)+\frac{R^2}{U^2}\left(\left(1+w'(\rho)^2\right)d\rho^2+\rho^2d\Omega_3^2\right)\, .
\end{equation}
We set $X^8=w(\rho)$ and $X^9=0$ from the rotational symmetry of $X^8-X^9$-plane.

\vspace{.3cm}
We briefly review of the typical two types of embedding of the D7 brane, which are
characterized by the profile function $w(\rho)$ given above. We denote it as
$w_0(\rho)$ for the case
of $F_{ab}=0$.
In this case, there is no gauge field ($F_{ab}=0$) and the on-shell action of D7-brane is obtained by 
substituting $w_0(\rho)$ as
\bea 
 S_{D7} &=& - \int d^4x L_{D7}\, , \label{D7action-full}  \\
   L_{D7} &=& 2\pi^2 \tau_7 \int d\rho {\cal L}_{D7}\, , \quad
          {\cal L}_{D7}=\frac{r^4}{U^4}\rho^3f(r)\sqrt{\left(1+w'_0(\rho)^2\right)}\, . \label{D7action-in}
\eea
We notice that $L_{D7}$, which is evaluated at on-shell with appropriate counterterms, is considered as the effective Lagrangian of the Yang-Mills theory with
a quark.

\begin{figure}[htbp]
\vspace{.3cm}
\begin{center}
\includegraphics[height=6cm,width=7cm]{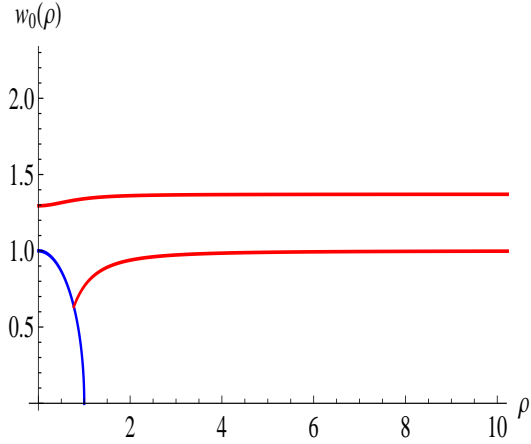}%
\caption{The typical embedding solutions  $w_0(\rho)$ are shown for $r_T=\sqrt{2}$ and $R=1$. 
The upper one is for $m_q = 1.37$ (Minkowski embedding), and the lower curve for $m_q=1$ 
(Black Hole embedding). \label{wsol}}
\end{center}
\end{figure}

The two typical solutions, which are called as the black hole (BH) embedding and the Minkowski embedding
respectively, 
of $w_0(\rho)$ are shown in the Fig. \ref{wsol}. 
Here we give the following comments on these solutions.

\begin{itemize}
\item At large $\rho$, $w_0(\rho)$ behaves as
\begin{equation}\label{w0}
w_0(\rho)=m_q+\frac{c}{\rho^2}+\cdots,
\end{equation}
we can get the value of the current quark mass $m_q$ and {the chiral condensate $c\equiv -\langle \bar{\Psi}\Psi\rangle $} respectively.
So the effective Lagrangian is given as a function of $m_q$, $T$, and $c$.
\item As shown in the Fig.\ref{wsol}, 
 in the Minkowski embedding solutions, the D7-brane is off the horizon $r_T$. 
On the other hand, the black hole embedding solutions attach to the horizon.

\end{itemize}
 
\section{\bf Holographic Schwinger Effect}
\vspace{.3cm}
\noindent{\bf External electric field}

We add non-trivial gauge field in the D7 action to impose an external electric field $E$ on the 
system considered in the previous section. We consider the following two cases with $E$.

\vspace{.3cm}
{\bf (A)} The first case is to find a stable state by imposing the external electric field $E$  \cite{KO}.
$E$ is imposed on this system through $A_a$, which is defined as
$F_{ab}=\partial_aA_b-\partial_bA_a$, as
\begin{equation}
 \tilde{A}_x\equiv 2\pi\alpha ' A_x=-Et+h(\rho), \label{E-field}  
\end{equation}
and the other components are zero.
In this case, ${\cal L}_{D7}$ is written as
\begin{equation}
{\cal L}_{D7}=\frac{r^2\rho^3}{U^4}\sqrt{f^2r^2U^2h'(\rho)^2+(f^2r^4-E^2R^4)(1+w'(\rho)^2))}\, , \label{ld7dbi}
\end{equation}
where $h'(\rho)=\partial_{\rho}h(\rho)$.
Then, the electric properties of the system under a constant electric field $E$
have been studied by solving the equations of motion
for $h(\rho)$ and $w(\rho)$
obtained from the above Lagrangian (\ref{ld7dbi}) \cite{KO,Oba,AFJK,Erd1,GIT}.
In this case, $h(\rho)$ is introduced to realize a non equilibrium steady state with an electric current $j$,
which is defined as
\begin{equation}
  {\partial{\cal L}_{D 7} \over \partial h'(\rho)}\equiv j.
\end{equation}

\vspace{.3cm}
{\bf (B)} The second case is to set $\tilde{A}_x$ as follows,
\begin{equation}
   \tilde{A}_x=-Et  ,  \label{E-field-2}  
\end{equation}
instead of (\ref{E-field}) \cite{Hashi}. 
In this case, the electric current is absent since we set as $h(\rho)=0$. 
This setting implies that the electric field suddenly turned on at $t=0$.
As a result, we find a false vacuum which induces an imaginary part of $L_{D7}$ 
given by (\ref{D7action-in}). This 
imaginary part, Im$L_{D7}$, is considered as the transition rate to a stable state with a
constant electric current $j$ as given in (A). Then this is 
connected to the production rate of the electric charges, which becomes the source
of this electric current $j$. This phenomenon is therefore related to
the Schwinger effect. In fact, in this case, (\ref{ld7dbi}) is written as
\begin{equation}
 {\cal L}_{D7}= \frac{R^2r^2}{U^4}\rho^3\sqrt{\left(1+w'_0(\rho)^2\right)\left(\frac{r^4f^2}{R^4}-E^2\right)} \label{ld7j0}\, ,
\end{equation}
where $w_0(\rho)$ denotes the solution of the equation of motion for $E=0$, just before imposing $E$. 
This is proportional to $\sqrt{\frac{r^4f^2}{R^4}-E^2}$, and then $ {\cal L}_{D7}$ becomes imaginary in an appropriate region of $\rho$ where $\frac{r^4f^2}{R^4} < E^2$ is satisfied. 
The lower bound of the electric field, $E=E_c$, is given as the minimum value of $r^4f^2/R ^4$.

Then the production rate,
$\Gamma$, of the quark-antiquark pair is obtained. It is defined as follows,
\begin{equation}\label{ImL}
\Gamma\equiv {Im { L_{D7}}\over 2\pi^2\tau_7}=\int_{\rho_{min}}^{\rho_c} d\rho 
 \frac{R^2r^2}{U^4}\rho^3 \sqrt{\left(1+w'_0(\rho)^2\right)\left(E^2-\frac{r^4f^2}{R^4}\right)}.
    \label{imaginary-a}
\end{equation}
The upper limit $\rho_c$ is determined by the equation, $\frac{r^4f^2}{R^4}\big|_{\rho=\rho_c}=E^2$.
It is given by solving the following equations,
\begin{equation}\label{profile}
\rho_c=\sqrt{U_c^2-w_0(\rho_c)^2}\, ,
\end{equation}
where
\begin{equation}
U_c=r_c\sqrt{\frac{1+f(r_c)}{2}}\, ,
\end{equation}
and $r_c$ is defined as 
\begin{equation}
r_c^2f(r_c)/R^2=E.
\end{equation}
As for $\rho_{min}$, $\rho_{min}=0$ for the Minkowski embedding, and for the black hole embedding  it is
given as the point where the D7 brane touches at the horizon, namely $\rho_{min}^2+w_0^2(\rho_{min})=r_T^2/2$.

\vspace{.3cm}
In the next, we explain how to calculate the above imaginary part.

\begin{itemize}
\item In (\ref{ld7j0}) and (\ref{imaginary-a}), we notice that the profile function,
$w_0(\rho)$, is the solution 
obtained before imposing the external $E$ field. Namely, 
it is given in the previous section by (\ref{w0}). 
\item On the other hand,  $w(\rho)$ in (\ref{ld7dbi}) of ({\bf A}) represents the solution which should be
obtained by solving the equation of motion derived from  (\ref{ld7dbi}).
So we notice $w(\rho)\neq w_0(\rho)$.
\item Then it may be reasonable to regard the above imaginary part (\ref{ImL}) as
the transition probability from a 
destabilized state, $\{w_0(\rho), j=0, E\}$, to an equilibrium steady state,
$\{w(\rho),  j\neq 0, E\}$, for a fixed $E(>E_c)$ \cite{Hashi}.
\item The constant current $j$ must be supported by some free charges. They should be generated
by the Schwinger effect from the vacuum state. In this sense, $\Gamma$ could be regarded as the 
production rate of quark and anti-quark pair. Further,  $L_{D7}$ could be regarded as the Euler-Heisenberg Lagrangian
for ${\cal N}=4$ SYM theory coupled to ${\cal N}=2$ hypermultiplet.
\end{itemize}

\vspace{.3cm}
As mentioned above, $m_q$ and $c$ are already included in the above Lagrangian  (\ref{ld7j0}).
To find the production rate,
the remaining work is to perform the integration of (\ref{imaginary-a}) with respect to
$\rho$ over an appropriate range. 

\vspace{.3cm}
\noindent{\bf $E$ dependence of $\Gamma$ and Physical Quantities}

\vspace{.3cm}
The above expression of $\Gamma$ contains various parameters,
$T$, $m_q$, $c$ and $E$. As for the $E$ dependence, it is absorbed in the dimensional
parameters in $\Gamma$, for example as $\tilde{r}=r/\sqrt{E}$.  
 The dimensionful factor $E^2$ is separated out as a prefactor to fix the dimension of $\Gamma$. Namely, 
we can
rewrite $\Gamma$ as, 
\bea
  { \Gamma \over 2\pi^2\tau_7} &\equiv & E^2 ~\gamma(\tilde{m}_q, \tilde{T})\, , 
                 \label{scale-1} \\
 \gamma(\tilde{m}_q, \tilde{T}) &=& \int_{\tilde{\rho}_{min}}^{\tilde{\rho}_c} d \tilde{\rho} 
 \frac{R^2\tilde{r}^2}{\tilde{U}^4}\tilde{\rho}^3 \sqrt{\left(1+\tilde{w}'_0(\tilde{\rho})^2\right)
      \left(1-\frac{\tilde{r}^4f^2}{R^4}\right)}\, , \label{scale-2}
\eea 
where $\tilde{m}_q=m_q/\sqrt{E}$, $\tilde{T}=T/\sqrt{E}$, and $f=\sqrt{1-(\tilde{r}_T/\tilde{r})^4
}$.

In this case, $\gamma$ is normalized as
\beq
  \gamma(0, 0)= {R^4\pi\over 8}\, .
\eeq

In the above expression (\ref{scale-1}), the dimensionful factor $E^2$ is factored out, and the remaining $E$s are
absorbed into $\gamma(\tilde{m}_q, \tilde{T})$ in terms of the newly defined variables, $\tilde{m}_q, \tilde{T}, $ etc. 
Then, the dependences on
$\tilde{m}_q$ and $\tilde{T}$ can be seen from $m_q$ and $T$ dependence of $\gamma$ for a fixed $E$.
So it is enough to examine $\Gamma$ for one value of $E$ to see the $m_q$ and $T$ dependence. 
We perform the analysis for
a fixed value of $E$ by varying other parameters, $m_q$ and $T$, hereafter. 
As for the $E$ dependence,
we discuss in the last section.

\section{Deconfining Chiral Symmetric Phase}

As shown above, $\Gamma$ depends on $w_0(\rho)$, which reflects the chiral symmetry
of the vacuum state before imposing the external $E$. At first, we examine the 
case of the chiral symmetric phase given by the metric (\ref{adsbh}). While, for $m_q>0$, $w_0$
is given in the form of (\ref{w0}) with negative $c$, $c=0$ is found for $m_q=0$.
Then we can say that the chiral symmetry is realized in this phase. 

\subsection{Quark Mass Dependence at $T=0$, AdS$_5\times S^5$ Limit}

\begin{figure}[htbp]
\vspace{.3cm}
\begin{center}
 \includegraphics[height=6cm,width=6.5cm]{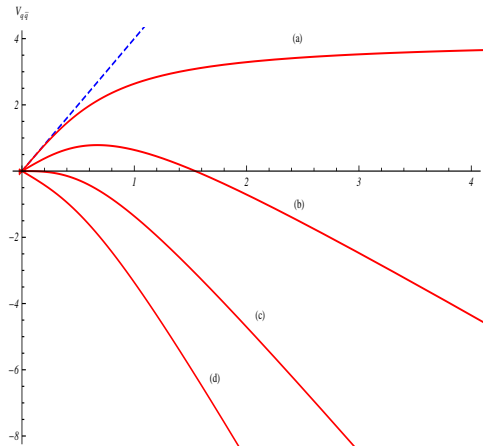}
 \caption{Quark-antiquark potentials $V_{q\bar{q}}$ defined in (\ref{qpot}) are shown as functions of 
the distance $l$ between quark and antiquark for  $m_q=2$, $R=1$ and various $E$. The curves (a), (b), (c) and (d) 
are for $E=0,~2,~4,~$ and 6 respectively. Here 
 $r_{max}=w_0(0)=m_q=2$ and  $E_c=m_q^2/R^2=4$. The dashed blue line represents the tangential line $m_q^2l/R^2$ at the origin for $V_{q\bar{q}}$ with $E=0$.
 \label{vads}}
\end{center}
\end{figure}

Before considering the metric (\ref{adsbh}) for arbitrary $T$, we study $m_q$ dependence at $T=0$, namely
the AdS$_5\times S^5$ limit. In this case, $w_0$ is given as a constant (and $c=0$),
\beq
    w_0=m_q\, ,
\eeq
for any $m_q$ as shown in \cite{KBS}. This state is supersymmetric, and
$\Gamma$ is expressed as
\beq\label{zeroT}
   \Gamma (m_q,T=0)\equiv \Gamma_0(m_q)\, ,
\eeq
where $E$ is also included in $\Gamma$, but it is abbreviated for simplicity.
Then, {(\ref{imaginary-a}) becomes}
\beq\label{zeroT-1}
  \Gamma_0(m_q)
=\int_0^{\rho^*}d\rho~ \rho^3\sqrt{{R^4E^2\over r^4}-1}\, , \quad
    \rho^*=\sqrt{R^2 E^2-m_q^2}\, . \label{T-0G}
\eeq
This is evaluated as
\beq\label{T0G}
  \Gamma_0 (m_q)={R^4E^2\over 2}I(\theta_0)\, ,
\eeq
where 
\beq
  I(\theta) ={\pi\over 4}-{\theta\over 2}+{1\over 2}\sin \theta\left(
        \cos \theta - \log\left({1+\cos \theta\over 1-\cos \theta }\right)\right)\, ,
\eeq
and
\beq
        \sin \theta_0 = {m_q^2\over R^2E}\left(={\tilde{m}_q^2\over R^2}\right)\, .
\eeq
We notice that there is a lower bound of $E$,
\beq\label{bound}
    E \geq {m_q^2\over R^2}\, (=E_c)\, ,
\eeq

to have a finite $\Gamma$ since $\sin \theta_0 \leq 1$ or $\rho^*$ should be real and finite.

To understand the above production rate, 
we consider the pair produced quark-antiquark total potential $V_{q\bar{q}}$ \cite{SZ} for various values of $E$.
This is defined as
\begin{equation} \label{qpot}
V_{q\bar{q}}=U_{q\bar{q}}-El
\end{equation}
where $U_{q\bar{q}}$ \footnote{$U_{q\bar{q}}$ and $l$ can be calculated from the on-shell Nambu-Goto action of the string whose endpoints are  on the D7-brane\cite{KMMW}, and they are given for AdS$_5\times S^5$ background as 
\begin{eqnarray}
U_{q\bar{q}} &=&  {1\over \pi\alpha '}\int_{r_{min}}^{r_{max}}dr\frac{1}{\sqrt{1-(r_{min}/r)^4}}, \nonumber \\
l &=& 2R^2\int_{r_{min}}^{r_{max}}dr\frac{1}{\sqrt{(r/r_{min})^4-1}}\, , \nonumber
\end{eqnarray}
where $r_{min}$ is the bottom point of the  string, and $r_{max}$ is the position of the string endpoints  on the D7-brane.}
is the quark-antiquark potential at a distance $l$ without the electric field $E$.
The potential $U_{q\bar{q}}$ is obtained from the action of a string whose endpoints  are on the D7-brane\cite{KMMW}. 
The endpoint, $r_{max}$, can be taken at the various position on the D7-brane and the lowest value of $r_{max}$ 
corresponds to $m_q$.   
In the Fig. \ref{vads}, we show the case of  $r_{max}=m_q$.
The total potential $V_{q\bar{q}}$ are shown for 
$E=0$ ((a)),  $0<E<E_c$ ((b)), $E=E_c$ ((c)), and $E>E_c$ ((d)) respectively.

\vspace{.3cm}
We could find from the potential for $E=0$  (curve (a) in the Fig.\ref{vads})
that the bound value, $m_q^2/ R^2$, of (\ref{bound}) corresponds to the tension of the quark 
and anti-quark potential $V_{q\bar{q}}$ observed at very
short distance as shown in the Fig.\ref{vads} by the dotted line. This tension is responsible 
for constructing a quark and anti-quark bound state, a meson\cite{KMMW}. 
At large distance, it behaves like a Coulomb potential and no linear potential is observed.
So (\ref{bound}) is a sufficient condition to remove the attractive force, which is responsible to make a bound state
of the pair produced quark and anti-quark, from $U_{q\bar{q}}$. In fact, for $E\ge E_c$, we find $V_{q\bar{q}}\leq 0$ 
for whole range of $l$ as shown by the curves (c) and (d). Thus, in the case of (\ref{bound}), no tunneling process
is needed to make a steady current.

\begin{figure}[htbp]
\vspace{.3cm}
\begin{center}
\includegraphics[height=6cm,width=15cm]{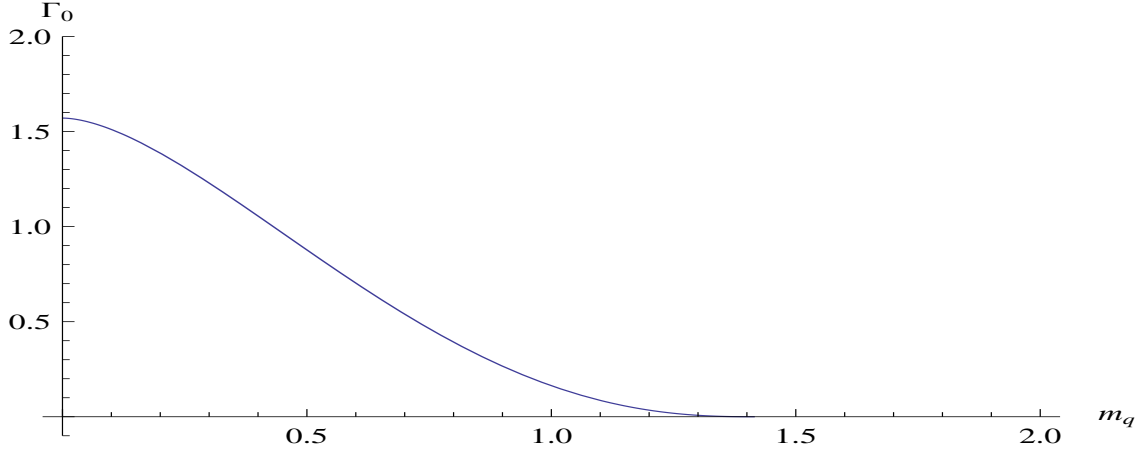}
 \caption{$\Gamma_0(m_q)$ is shown for $E=2$ and $R=1$.  The curve  
represnts the formula (\ref{T-0G}). 
\label{mass-dep-2}}
\end{center}
\end{figure}

\vspace{.3cm}
 As a result, we could say that $\Gamma_0(m_q)$ given by (\ref{T-0G}) represents the probability of the transition 
from a false vacuum with
$j=0$ to an equilibrium steady state with a finite $j$ for a given $E (\geq {m_q^2\over R^2})$. 
The latter state is obtained according to the way given in ({\bf A}) in the previous section.
Then, our result, the formula (\ref{T0G}), is shown 
in the Fig.\ref{mass-dep-2} for $E=2$. 

 In the Fig.\ref{mass-dep-2}, non-zero $\Gamma$ is obtained for $m_q\leq\sqrt{E}R(=\sqrt{2})$.
For the case of $m_q > \sqrt{E}R$, the vacuum decay does not occur since the imposed $E$ is less than the critical value $E_c$.
Namely, for $0<E<E_c$, the attractive potential remains as shown by the curve (b)
in the Fig. \ref{vads}. 
Then in this case, a tunneling process
would be needed to produce free quark pairs. 
The tunneling process would be important in this case. 
We will give its analysis in a future work. Here, the analysis is restricted to the case
of (c) and (d) of the  Fig.\ref{vads}.

\vspace{.3cm}
\subsection{Temperature Dependence}

Now we study the temperature dependence of the production rate, which is denoted as
$$\Gamma(T, m_q)\equiv\Gamma_T(m_q).$$ 

\begin{figure}[htbp]
\vspace{.3cm}
\begin{center}
 \includegraphics[height=6cm,width=6cm]{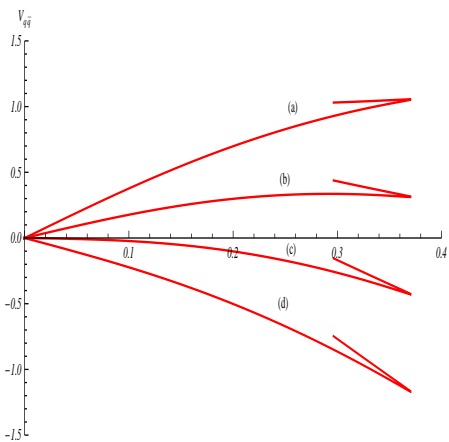}
  \caption{Quark-antiquark potential $V_{q\bar{q}}$ at finite temperature with $T=1/\pi$ is calculated as a function of  the distance $l$ between quarks  for $E=0,~2,~4,~6$ as (a),~(b),~(c) and (d)  respectively, where $R=1$, $r_{max}=w_0(0)=2$ and   
 $E_c=4$.\label{vqqft-fig}}
\end{center}
\end{figure}
At finite $T$, we could see the similar behavior of $V_{q\bar{q}}$ to the one of the $T=0$ case given 
above.
In the case of Minkowski embedding, 
the quark-antiquark potential $V_{q\bar{q}}$ as shown in  Fig.\ref{vqqft-fig} disappears at the finite distance $l$  by the thermal screening. 
 Except for this point, the behavior of $V_{q\bar{q}}$ at small $l$ is similar to the case of $T=0$.
As for the case of BH embedding, $V_{q\bar{q}}$ behaves as the curves (c) and (d) in  Fig.\ref{vqqft-fig} since $E_c=0$ as shown below.

\vspace{.3cm}
\noindent{\bf Case of $m_q=0$
}
\vspace{.3cm}

At first, we consider the case of $m_q=0$. In the present case, the system is in a phase where
the chiral symmetry is restored. In fact, $w=0$ then $c=0$ at any $T$. 
 This implies that $\Gamma$ would be independent of $T$. 
We show this point below. 

For $m_q=0$, $w=w'=0$, $c=0$ and
\beq
  U=\rho\, , \quad r=\rho\sqrt{1+{r_T^4\over 4\rho^4}}\, .
\eeq
Then we obtain
\beq
   \Gamma_T(0)=R^2 \int_{\rho_{min}}^{\rho_{max}}{d\rho\over\rho}r^2\sqrt{E^2-{r^4f^2\over R^4}}\, ,
\eeq

where $\rho_{min}=r_T/\sqrt{2}$ and 
$\rho_{max}=\left\{ER^2\left(1+\sqrt{1+4\rho_{min}^4/(ER^2)^2}\right)/2\right\}^{1/2}$.
We can perform the above integration, and we arrive at
 \beq\label{m0g}
   \Gamma_T(0)={\pi R^4\over 8}E^2\, .
\eeq

It is noticed that the above result is independent of $T$.
 In fact (\ref{m0g}) is the same value as (\ref{T0G}) with $m_q=0$,  
\beq\label{t-inf}
     \Gamma_0(0)={R^4E^2\over 2}I(0)={\pi R^4\over 8}E^2\, .
\eeq 

\begin{figure}[htbp]
\vspace{.3cm}
\begin{center}
 \includegraphics[height=6cm,width=6cm]{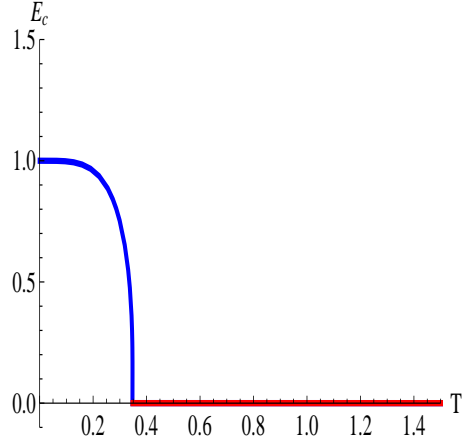}
  \caption{The relation between  $E_c$ and $T$ for $m_q=1$ with $R=1$. The blue curve is obtained by Minkowski embedding solutions and red line is given by BH embedding solutions. \label{Ec-T}}
\end{center}
\end{figure}


\vspace{.3cm}
\noindent{\bf Case of $m_q>0$} 
\vspace{.3cm}

In this case, the lower bound $E_c$ exists. An example is shown
in the Fig.\ref{Ec-T} for $m_q=1$ with $R=1$. As seen from the Fig.\ref{Ec-T},  $E_c$ changes drastically 
near $T_c\sim 0.35$,
where the type of embedding changes from Minkowski to BH type.
For $T>T_c$, BH embedding is realized and there is no stable
meson in this phase since the attractive force at short distance disappears due to the thermal screening. 
Then we find $E_c=0$ for $T>T_c$.
The transition temperature, $T_c$, depends on the current quark mass $m_q$.

\vspace{.3cm}
\noindent{\bf Numerical Results of $\Gamma$}

\begin{figure}[htbp]
\vspace{.3cm}
\begin{center}
 \includegraphics[height=6cm,width=6.5cm]{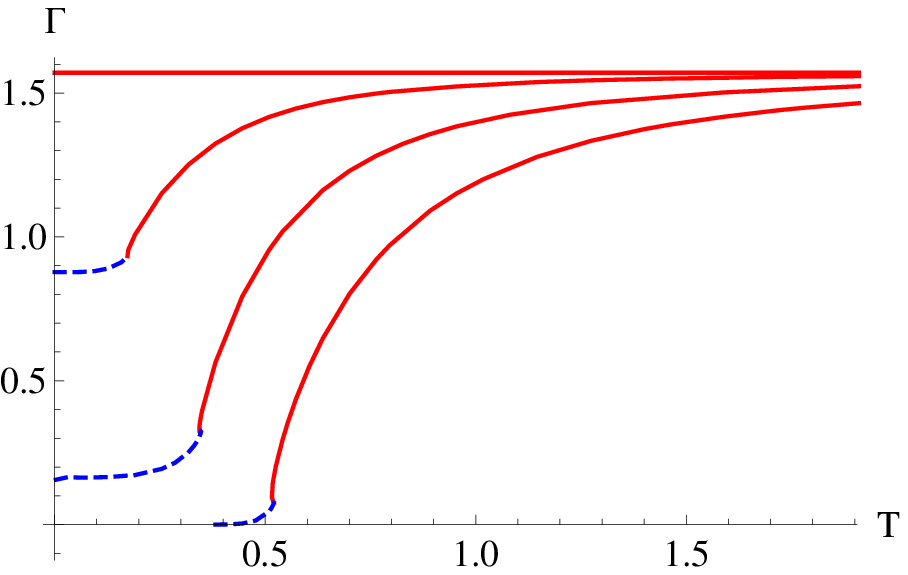}
  \includegraphics[height=6cm,width=6.5cm]{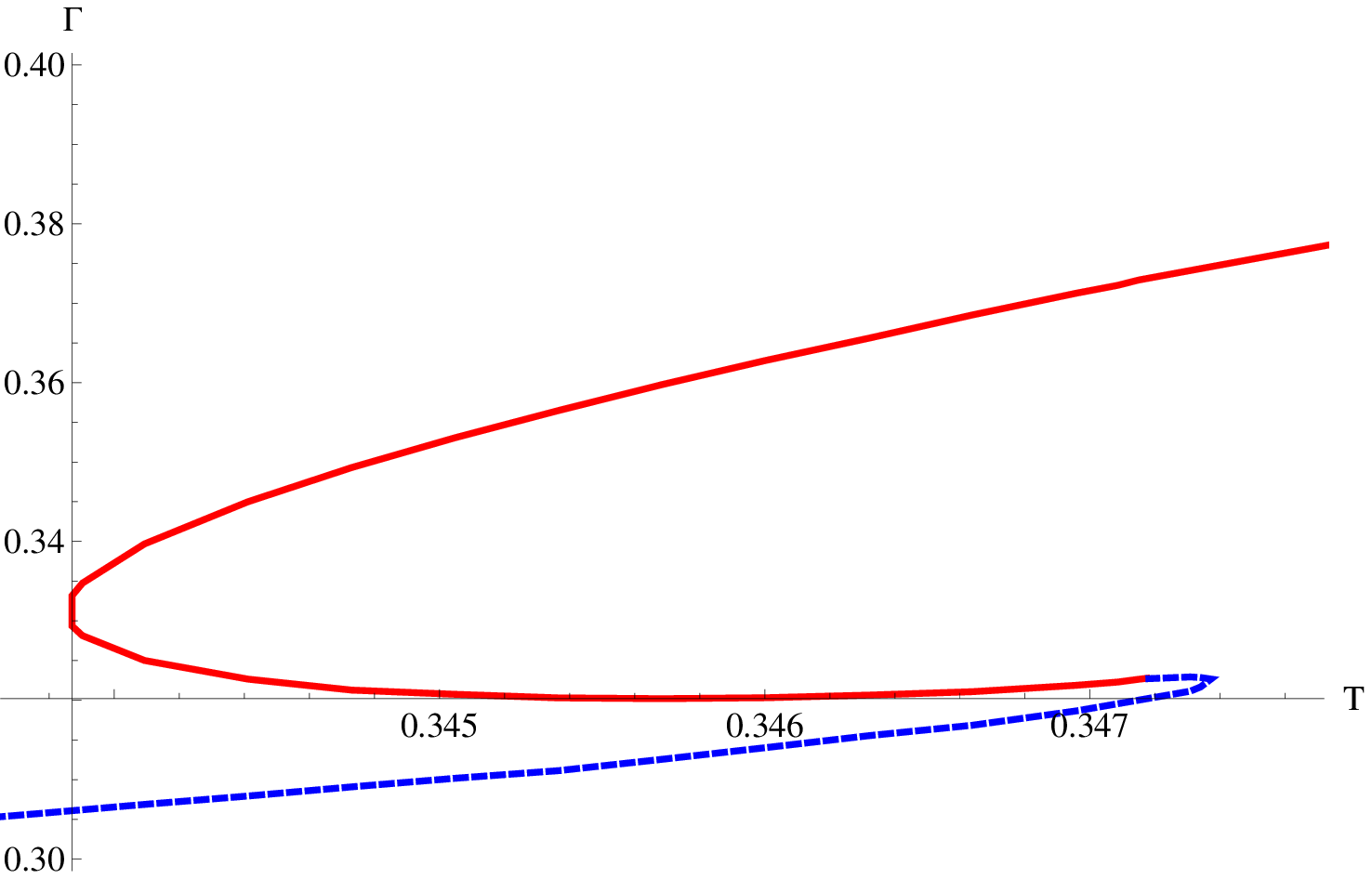}
\caption[width=10cm]{The left figure shows  the relation between $\Gamma$ and $T$ with $E=2$ and $m_q=0,~0.5,~1,~1.5$ from the above. The dashed blue curves are given by the Minkowski embedding solutions and the solid red curves are obtained by the BH embedding solutions. 
For $m_q=1.5$, $\Gamma$ has a finite value  for  $T \ge 0.38$. 
The right figure shows the  extended region near the transition point with  $m_q=1$. We can see the first order phase transition between Minkowski embedding and BH embedding.\label{gm}}
\end{center}
\end{figure}

\vspace{.3cm}
 In the  Fig.\ref{gm}, four results of numerical estimation of 
 $\Gamma_T(m_q)$ 
are shown for $m_q=1.5,~1.0,~0.5,~0$ with $E=2$. 

From this Fig.\ref{gm}, we find a first order transition of $\Gamma$ at $T=0.5,~0.34$ 
and $0.22$ for the cases of $m_q=1.5,~1.0$ and $0.5$ respectively. 
The transition temperature depends on the value of $m_q$. At this point, the embedding form
of the D7 brane transits from the ``Minkowski embedding'' to the  ``BH embedding". This is a well known
phase transition observed in the holographic SYM theory. Near this point,
$\Gamma$ increases rapidly and approaches to the high temperature limit more slowly at large $T$. 

 Fig.\ref{gm} also shows that $\Gamma_T(0)$ is independent of $T$  as shown in  (\ref{m0g}),  and $\Gamma_T(m_q)$ at  $m_q=1.5,~1.0,~0.5$ approach to $\Gamma_T(0)$ at $T\to \infty$. We will confirm this point with the analytic calculation as follows. 
\vspace{.3cm}

We evaluate the upper limit, 
$\Gamma_T(m_q)|_{T\to\infty}\equiv \Gamma_{\infty}(m_q)$.
At large $T$, $\Gamma_T(m_q)$ is approximated as 
\bea
     \Gamma_{\infty} (m_q) &=& \int_1^{x^*}~dx~F(x)\, , \\
      F(x)&=& \left({U_TR\over 2}\right)^2{1+x\over x\sqrt{x}}\sqrt{E^2-\left({U_T\over R}\right)^4 f(x)}\, , \\
      f(x) &=& x\left(1-{1\over x}\right)^2\, , \quad x=\left({U\over U_T}\right)^4\, ,
 \label{large-T}
\eea
where $w'(\rho)$ is approximated as being negligible small and
\bea
   U_T&=& {r_T\over \sqrt{2}}\, , \quad x^*=\alpha \left(1+\sqrt{1-{1\over\alpha^2}}\right)\, , \\
   \alpha &=&1+{R^4E^2\over 2U_T^4}\, .  \label{large-T-2}
\eea
 Then we find 
\beq \label{tinfg}
     \Gamma_{T} (m_q)|_{T\to \infty} = \left.{R^4E^2\pi\over 8}\left(1-{4R^2\over 3\pi U_T^2}E+O(1/U_T^4)\right)
      \right|_{T\to\infty}
                 =  {\pi R^4\over 8}E^2\, .
\eeq
This result is independent of $m_q$ and equivalent to $\Gamma_0(0)$ as shown in (\ref{t-inf}). The limiting value of $\Gamma$ for all $m_q$ approaches to this value at $T\to\infty$ as shown in Fig.\ref{gm}.


\begin{figure}[htbp]
\vspace{.3cm}
\begin{center}
 \includegraphics[height=6cm,width=6.5cm]{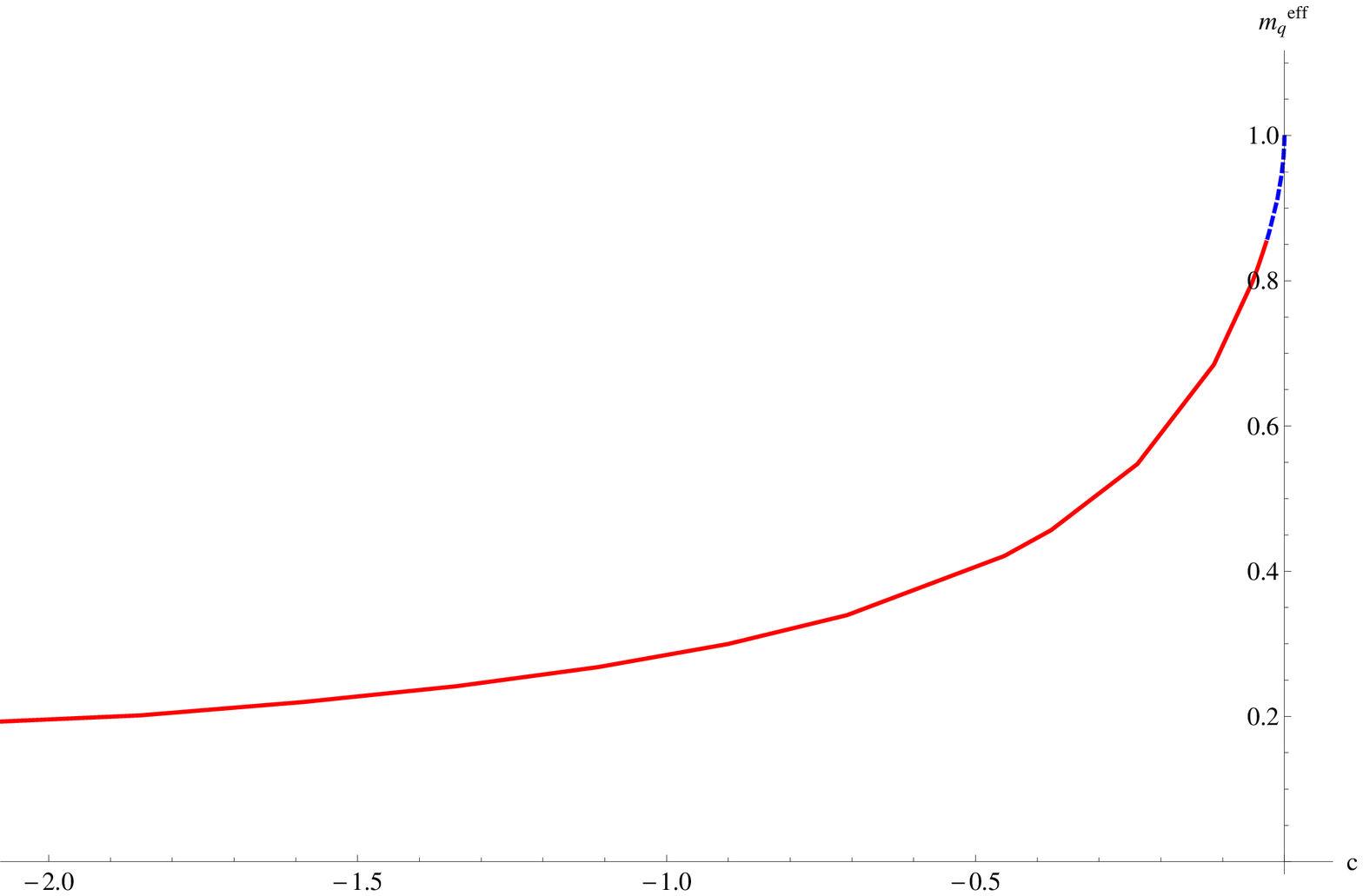}
 \includegraphics[height=6cm,width=7cm]{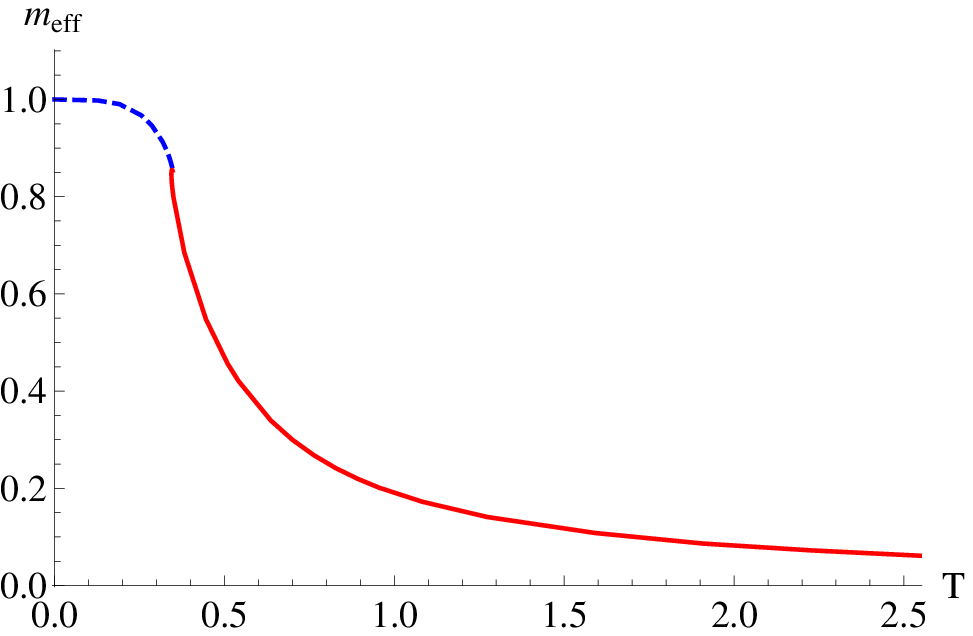}
 \caption{The $c$ (Left) and $T$ (Right) dependences of 
$m_q^{eff}$ for ${m}_q=1$, $E=2$ and $R=1$. The dashed blue curve is obtained by Minkowski embedding solutions and solid red line is given by BH embedding solutions. \label{T-meff}}
\end{center}
\end{figure}


\vspace{.3cm}
\noindent{\bf  Effective Mass $m_q^{\rm eff}$}

\vspace{.3cm}
As mentioned above, in the present case, we could find the chiral condensate $c$, which is negative, $c < 0$.
Its value depends on $m_q$ and $T$. 
Then, $c$ decreases from $c=0$ at $T=0$ with increasing $T$ monotonically. When we remember a simple
NJL formula, (\ref{NJL-c}) with $f(c)=1$, we may expect
that the effective quark mass $m_q^{eff}$ would be suppressed from the current quark mass $m_q$,
namely $m_q^{eff} < m_q$. Here we notice that the sign of $c$ is opposite to 
the chiral symmetry broken phase. 
It may be obtained after a calculation of the
self-energy with full quantum correction. It is however difficult to derive
$m_q^{eff}$ at finite temperature from the 4D SYM field theory side.

\vspace{.3cm}
We propose a way to estimate $m_q^{eff}$ by using an ansatz that
the dynamical effects of the temperature are absorbed into $m_q^{eff}$ and the production rate
is replaced by the one given for the AdS$_5\times S^5$ as given below. 

This ansatz is based on the following speculation. For the case of AdS$_5\times S^5$, the mass
of the pair created quark 
receives no correction from 
the gauge interaction due to the preserved supersymmetry. 
For non-supersymmetric cases studied here for the finite temperature phase (and for the confinement phase in the below), on the other hand, the 
quark mass is modified to $m_q^{eff}$ by the correction. 
The quark potential 
is similar to the case of AdS$_5\times S^5$ due to the imposed $E(>E_c)$. So the difference is reduced to the quark mass. According to this consideration, the value of $m_q^{eff}$ is obtained by comparing $\Gamma$ for the non-conformal case with the one
for the conformal case  at $T=0$.

Thus, under this ansatz,
the production rate $\Gamma$ at $T(\neq 0)$ can be related to $\Gamma$ at $T=0$ ,
which is obtained for AdS$_5\times S^5$ bulk, 
as follows,
\beq\label{meff}
    \Gamma(m_q,T)\equiv \Gamma_T(m_q)=\Gamma_0 (m_q^{eff})\, ,
\eeq
where  $m_q^{eff}$ is a function of $m_q$ and $T$ (or $c$). 
 As shown in the Fig. \ref{gm}, the T-dependence of $\Gamma_T(m_q)$ for fixed $m_q$ is read as 
\beq\label{eq-ineq}
   \Gamma_0 (m_q)\leq  \Gamma_T(m_q)\leq   \Gamma_{\infty}(m_q)\, , 
\eeq
and from (\ref{t-inf}) and (\ref{tinfg}) we also find
\beq\label{eq-eq}
   \Gamma_{\infty}(m_q) =\Gamma_0(0)\, .
\eeq
Namely, the value of $ \Gamma_T(m_q)$ is bounded between $ \Gamma_0 (m_q)$ and $\Gamma_0(0)$.
Therefore, in the whole range of the temperature, $0\leq T \leq \infty$, the value of $m_q^{eff}$
could be found between $m_q$ and zero by using the relation (\ref{meff}).
 While the equality  (\ref{eq-eq}) can be shown analytically by (\ref{t-inf}),
the inequality (\ref{eq-ineq}) on the other hand
assured by the numerical calculation as shown in the Fig. \ref{gm}.

\vspace{.3cm}
 Here we give a comment on the relation of our method to determine $m_q^{eff}$
and a way to use the on-shell action of 
a string which connects the D7 brane and the horizon at finite temperature \cite{MMT}. The result of the latter method
is in general different from ours. This discrepancy is clear in the case of BH embedding, where the method 
of the string action leads to $m_q^{eff}=0$. In our case, however, it is finite, and $m_q^{eff}=0$ is found only at the limit of the infinite temperature.

The reason of this disagreement is in the fact that, in our calculation of $\Gamma_T(m_q)$, 
various positions of $r$ on the D7 brane, 
where the pair production occurs, are taken into account of. 
In other words, $\Gamma$ is obtained by integrating
over $\rho$ from $\rho_{min}$ to $\rho_c$. The effective quark mass depends on the position of $\rho$. 
Then,
after an average of these various positions for the effective quark mass, we have arrived at our result.
At $T=\infty$, $m_q^{eff}=0$ is obtained since the position is restricted to the one on the horizon.

\vspace{.3cm}
\noindent{\bf  T and $c$ dependence of $m_q^{\rm eff}$}

\vspace{.3cm}
Then the $c$ and $T$ dependences of $m_q^{\rm eff}$ are obtained by using (\ref{meff}). 
The results are shown for $m_q=1$ and
$E=2$ in the Fig.\ref{T-meff}. As expected, the effective mass
decreases with $T$ rapidly near the transition point
($T_c\sim 0.35$) and it slowly decreases in the region of large $T$.
Then $m_q^{\rm eff}= 0$ is realized in the limit of $T=\infty$. 

\vspace{.3cm}
At the same time, $m_q^{\rm eff}$ is also plotted as a function of the chiral condensate $c$,
and it is shown in the left figure of the Fig. \ref{T-meff}. Near the transition point from
the Minkowski to the BH type embedding, it changes very rapidly. 

\subsection{\bf $m_q^{eff}$ and NJL coupling}


\vspace{.3cm}
In general, the effective mass of the quark is 
intimately related to the chiral condensate $-\langle\bar{\Psi} \Psi\rangle\equiv c$
and then it could be expressed as a function of $c$. It
is important
to find a precise form of the effective mass as $m_q^{eff}(c)$ for a fixed $m_q$.
This kind of analysis is usually performed in terms of
the Nambu-Jona-Lasinio (NJL) model \cite{NJ} for QCD, where $m_q^{eff}(c)$ is obtained
by supposing the effective Lagrangian of the quark with multi quark coupling terms.

In the present case, we are considering the ${\cal N}=2$ SYM theory. At $T=0$, 
we find $c=0$ in this theory
due to the supersymmetry. On the other hand, at finite temperature, we find negative $c$.
The problem is to see how our results are understood from   
 a NJL model for the ${\cal N}=2$ SYM theory
at finite temperature.  This is simply an extension of NJL to the high temperature phase.

\begin{figure}[htbp]
\vspace{.3cm}
\begin{center}
 \includegraphics[height=6cm,width=8.0cm]{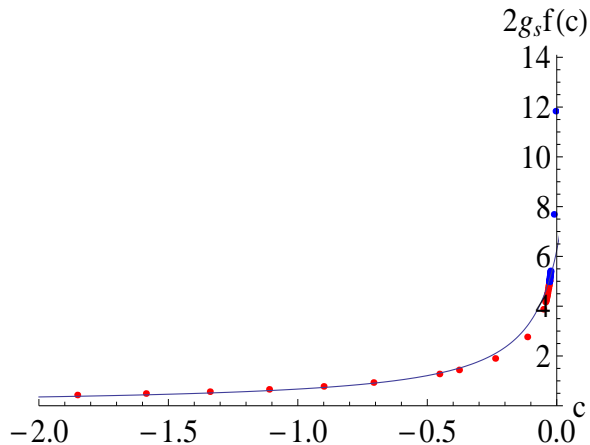}
  \caption{The relation between $2g_sf(c)$ in (\ref{NJL-c}) and $c$ for $m_q=1$, $E=2$ and  $R=1$ for the finite temperature case.
The solid curve represents $2g_sf(c)=6.25/(1-8.3 c)$.
\label{FT-gs} }
\end{center}
\end{figure}

\vspace{.5cm}

The NJL model is set as follows \cite{Kle},
\beq
  L_{NJL}=\bar{\Psi}\left(i\gamma^{\mu}\partial_{\mu}-m_q\right)\Psi+g_{s}(\bar{\Psi}\Psi)^2
+\cdots\, .
\eeq
The details of this model are not discussed. We consider the simplest case of this model
to compare with our holographic results. 

 While it would be possible to consider the various types of condensates of the fermi field, 
we do not
consider them here since they would not contribute to the quark mass. 
Then, in the present analysis, 
the terms like $(\bar{\Psi}i\gamma_5\Psi)^2$ are neglected. 
By considering the mean field approximation and taking into account of the higher order terms with respect to
$\bar{\Psi}\Psi$,
the effective quark mass would be obtained in the form,
\beq\label{NJL-c}
   m_q^{eff} = m_q+2g_s c f(c) \,, \label{NJL-c}
\eeq
where $g_s$ is taken as a constant, and 
$f(c)$ denotes a function of $c$. In this case the 
higher order terms of $c$ are absorbed in $f(c)$. 
For the case of $m_q=1$ and $E=2$, $f(c)$ is shown in the Fig. \ref{FT-gs}, 
and we have a rough estimation,
\beq
   2g_sf(c)={6.25\over 1-8.3 c}\, .
\eeq
This implies that we need infinite series of 
$c$ ($1/c$) to explain this behavior at small (large)
$c$.
In any case, our result  would give an important clue to understand the dynamics of the theory
at finite temperature.

\vspace{.1cm}
\section{Confining phase}

We consider a holographic theory which is in a chiral symmetry broken and quark
confining phase. This is realized by adding a non-trivial dilaton, which corresponds to 
the vacuum with gauge condensate $\langle F_{\mu\nu} F^{\mu\nu}\rangle$ parametrized by 
$r_0^4$.
The bulk background, dual to confining gauge theory considered here, is expressed as \cite{GIT},
$$ 
ds^2_{10}=G_{MN}dX^{M}dX^{N} ~~~~~~~~~~~~~~\qquad
$$ 
\beq\label{background}
=e^{\Phi/2}
\left\{
{r^2 \over R^2}A^2(r)\left(-dt^2+(dx^i)^2\right)+
\frac{R^2}{r^2} dr^2+R^2 d\Omega_5^2 \right\} \ . 
\label{finite-c-sol}
\eeq
in the string frame. {$A(r)$ and the dilaton ${\Phi}$ are} given by
\begin{equation}\label{non-susy-sol}
A(r)=\left((1-(\frac{r_0}{r})^8)\right)^{1/4},\qquad
e^{\Phi}=\left(\frac{(r/r_0)^4+1}{(r/r_0)^4-1}\right)^{\sqrt{3/2}}\,\qquad ,
\end{equation}
respectively.
We should notice that this configuration has a singularity at the horizon $r=r_0$. 
So we can not extend our analysis to near this point. This difficulty would be
resolved by introducing higher curvature contributions.

Fortunately, all the embedding solutions used here
avoid the singularity for any region of the parameter which we used. 
This would be reasonable since a finite solution can not be defined at any singular point of the background.

\vspace{.3cm}
The extra six dimensional part of the above metric (\ref{finite-c-sol})
is rewritten as,
\beq
 \frac{R^2}{r^2} dr^2+R^2 d\Omega_5^2
 =\frac{R^2}{r^2}\left(d\rho^2+\rho^2d\Omega_3^2+(dX^8)^2+(dX^9)^2
\right)\ ,
\eeq
where $r^2=\rho^2+(X^8)^2+(X^9)^2$.
Then we obtain the induced metric for D7 brane,
\bea
ds^2_8 &=& e^{\Phi/2}
\left\{
{r^2 \over R^2}A^2\left(-dt^2+(dx^i)^2\right)+\right.
\left.\frac{R^2}{r^2}\left((1+(\partial_{\rho}w)^2)d\rho^2+\rho^2d\Omega_3^2\right)
 \right\} \,  \label{D7-metric} \\
  &=& e^{\Phi/2}\tilde{G}_{ab}d\xi^a d\xi^b\, ,
\eea
where we take as $X^8=w(\rho)$ and $X^9=0$. The suffices $a$ and $b$ run from 0 to 7.

\vspace{.3cm}
By taking the gauge field as $ \tilde{A}_x (\rho,t)=-Et+h(\rho)$, we arrive at the following D7 brane
action \cite{GIT},
\beq
S_{\rm D7} =-2\pi^2\tau_7~\int d^4x d{\rho}\rho^3 {R\over r}
   e^{\Phi/2}\sqrt{P e^{\Phi}-Q}
\ ,
\label{D7-action-2}
\eeq
\beq
 P=|\tilde{G}_{00}|\tilde{G}_{xx}\tilde{G}_{\rho\rho}\, ,
 \quad Q=\tilde{G}_{\rho\rho}{\dot{\tilde{A}}_x}^2
-|\tilde{G}_{00}|{{\tilde{A'}}_x}^2\, ,
\eeq
where $\tilde{G}_{MN}=e^{-\Phi/2}{G}_{MN}$. 
At first, we solve the equation of motion of ${\tilde{A'}}_x={h'}$ as
\beq\label{electric}
  e^{\Phi/2}{\rho^3\over r}{|\tilde{G}_{00}|\tilde{A'}_x
\over \sqrt{P e^{\Phi}-Q}}=J\, ,
\eeq
where $J$ denotes a constant and it corresponds to the electric current,
\beq
   J= \langle J_x\rangle \, .
\eeq
As for the solutions with and without $J$, they are seen in \cite{GIT}. 
The situation is similar to the above finite temperature case.
By setting $J$ as an appropriate value, we find a stable solution for $w(\rho)$ for a given $E$.

\begin{figure}[htbp]
\vspace{.3cm}
\begin{center}
 \includegraphics[height=6cm,width=6.5cm]{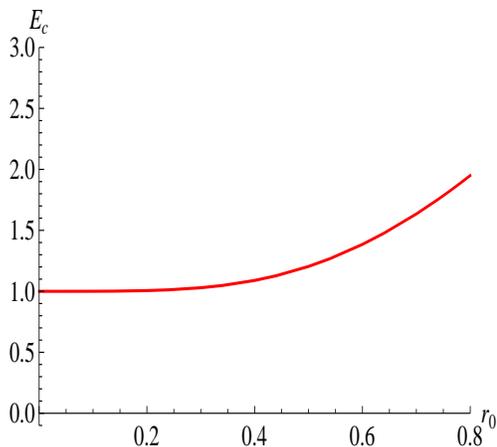}
  \caption{Relation between the critical electric field $E_c$ and $r_0$ for $m_q=1$ with $R=1$,\label{Ecr0}}
\end{center}
\end{figure}

\vspace{.3cm}
As in the case of the finite temperature theory, the imaginary part of the D7 action for a finite $E(\geq E_c)$, 
is given by setting $J=0$ as
\begin{equation}\label{non-S-im}
\Gamma={\rm{Im}~{\cal L}\over 2\pi^2\tau_7}=~\int_0^{\rho_c} d{\rho}
     \rho^3 A^2 e^{\Phi/2}\left({R\over r}\right)^2
          \sqrt{ 1+w_0'(\rho)^2}\sqrt{E^2-A^4 e^{\Phi}\left({R\over r}\right)^4}\, ,
\end{equation}
where 
\beq
 \rho_c=\sqrt{r_c^2-w_0^2(0)}\, , 
\eeq
and $r_c$ is defined as
\beq
    \left. A^4 e^{\Phi}\left({R\over r}\right)^4 \right|_{r=r_c}=E^2\, .
\eeq
where $E\geq E_c$ and the $r_0$ dependence of $E_c$ is shown in the Fig. \ref{Ecr0}.
 
\begin{figure}[htbp]
\vspace{.3cm}
\begin{center}
 \includegraphics[height=6cm,width=6.5cm]{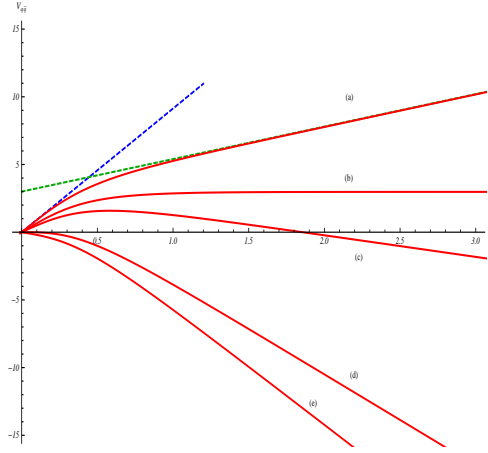}
  \caption{Quark-antiquark potentials $V_{q\bar{q}}$ are calculated as a function of  the distance $l$ between quarks  with $r_0=R=1$ and $r_{max}=w_0(0)=3$  for $E=0,~2.4,~4,~9.14,~11$ as (a),~(b),~(c),~(d) and (e) respectively. Here 
 $\tilde{\tau}_{SD}=E_c=9.14$ and $\tilde{\tau}_{QCD}=2.4$. The dashed blue line and dashed green line represent the tangential line $\tilde{\tau}_{SD}l$  at the origin and     $\tilde{\tau}_{QCD}l+1.5$ at the large $l$  for $V_{q\bar{q}}$ with $E=0$ respectively. 
 \label{vqq}}
\end{center}
\end{figure}

\vspace{.3cm}
Furthermore we notice the following points.
\begin{itemize}
\item The profile function $w_0(\rho)$ used in the Eq. (\ref{non-S-im}) is the one obtained 
from the D7 brane embedded before imposing the external electric field $E$. 
The equation of motion for $w_0(\rho)$ is therefore  obtained from the following D7 Lagrangian,
\beq\label{profile-eq0}
  {\cal L}_{D7}^{(0)}=-2\pi^2\tau_7\int d{\rho}\rho^3 A^4 e^{\Phi}\left({R\over r}\right)^4
          \sqrt{ 1+w_0'(\rho)^2}\, .
\eeq

\vspace{.3cm}
\item The second point is that there is a lower bound of $E$ in order to
have the above imaginary part. It is given as
\beq\label{non-susy-c}
  E^2\geq \left.{r^4A^4e^{\Phi}\over R^4}\right|_{r=r_c}
     \geq \left.{r^4A^4e^{\Phi}\over R^4}\right|_{r=w_0(0)}\equiv E_c^2.
\eeq
Notice  that there is no BH embedding in the present confining phase. Then the infrared end point of $w_0(\rho)$
is given by $w_0(0)$. 
\end{itemize}


In order to make free quark pair, it is necessary
to overcome the confining force $\tau_{QCD}$, which is the tension at large distance, by the external electric force $E$.
In the present model, $\tau_{QCD}$ is given by 
\beq
   2\pi\alpha'\tau_{QCD}\equiv \tilde{\tau}_{QCD}=\left.{r^2A^2e^{\Phi/2}\over R^2}\right|_{r=r*}\, ,
\eeq
where $r^*$ denotes the minimum point of $r^2A^2e^{\Phi/2}(r)$.

This is compared to the tension at
short distance, $\tau_{SD}$, which is given by
\beq
    2\pi\alpha'\tau_{SD}\equiv\tilde{\tau}_{SD}=\left.{r^2A^2e^{\Phi/2}\over  R^2}\right|_{r=w_0(0)}\, ,
\eeq
and $E_c$ is equivalent to $\tilde{\tau}_{SD}$ of the quark-antiquark potential calculated by the string whose endpoints are at $r_{max}=w(0)$ from (\ref{non-susy-c}).
Noticing $r* < w_0(0)$, we find 
\beq \label{tauqcd}
\tau_{QCD}<\tau_{SD} \, .
\eeq
Then we find that (\ref{non-susy-c}) is sufficient to remove 
the binding force between the pair produced quark and anti-quark. Then the tunneling process is absent. 
This point is explained below in terms of the quark-antiquark potential $V_{q\bar{q}}$ given by (\ref{qpot})
.

In confining case, $V_{q\bar{q}}$ is described by several curves, $(a)\sim (e)$, in  Fig.\ref{vqq}. 
 Here  (a), (b), (c), (d), (e) denotes the $V_{q\bar{q}}$ with  various E which satisfies $E=0$, $E=\tilde{\tau}_{QCD}$,  $\tilde{\tau}_{QCD}<E<\tilde{\tau}_{SD}$, $E=\tilde{\tau}_{SD}$ and $E>\tilde{\tau}_{SD}$ respectively.  
For $0\le E\le\tilde{\tau}_{SD}$ ((a) and (b)), 
we find $V_{q\bar{q}}>0$ for all region of $l$. Thus, we cannot find the free quark pair in this case.  For $\tilde{\tau}_{QCD}< E< \tilde{\tau}_{SD}$ ((c)), free quarks can be produced by the tunneling process.
For $\tilde{\tau}_{SD}=E_c\le E$ ((d) and (e)), $V_{q\bar{q}}$ becomes negative in all region of $l$. Thus, for (\ref{non-susy-c}), free quarks are produced without tunneling process.


\begin{figure}[htbp]
\vspace{.3cm}
\begin{center}
 \includegraphics[height=6cm,width=4.5cm]{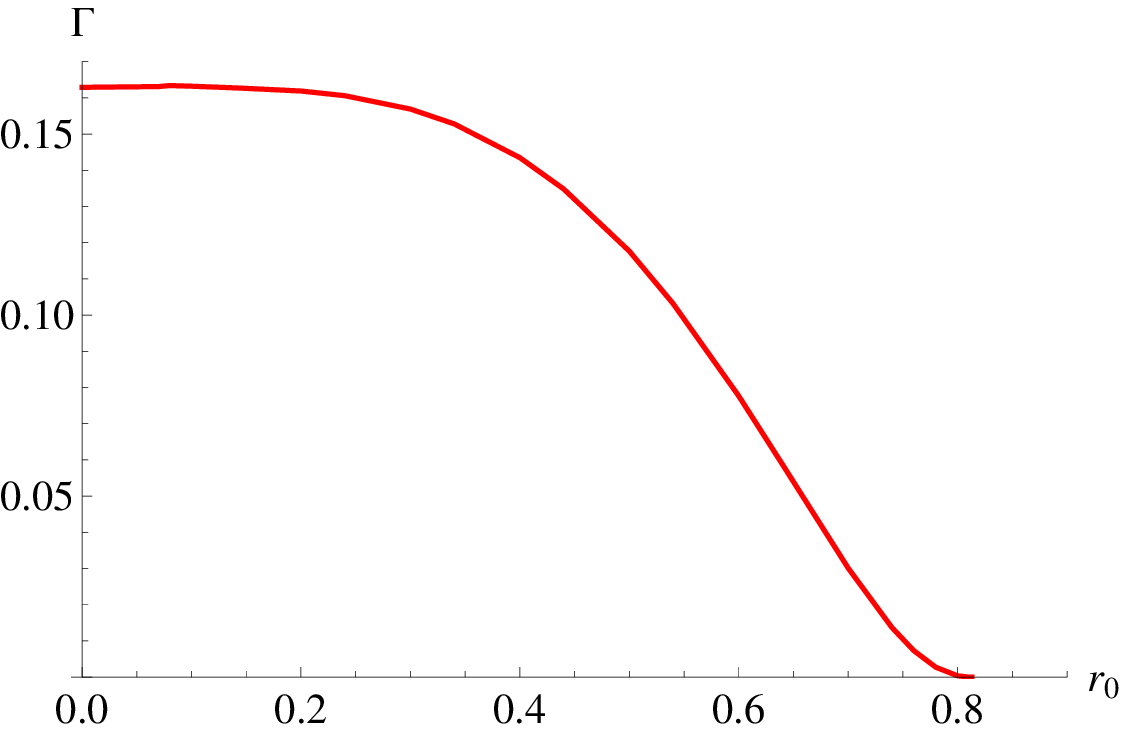}
 \includegraphics[height=6cm,width=4.5cm]{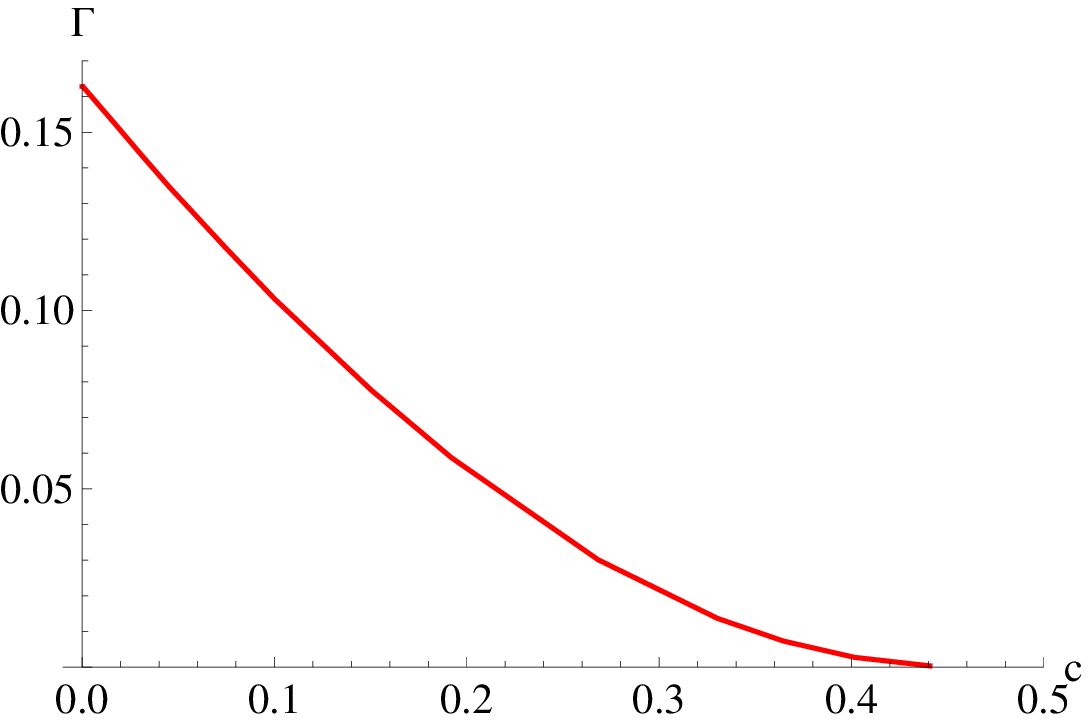}
  \includegraphics[height=6cm,width=4.5cm]{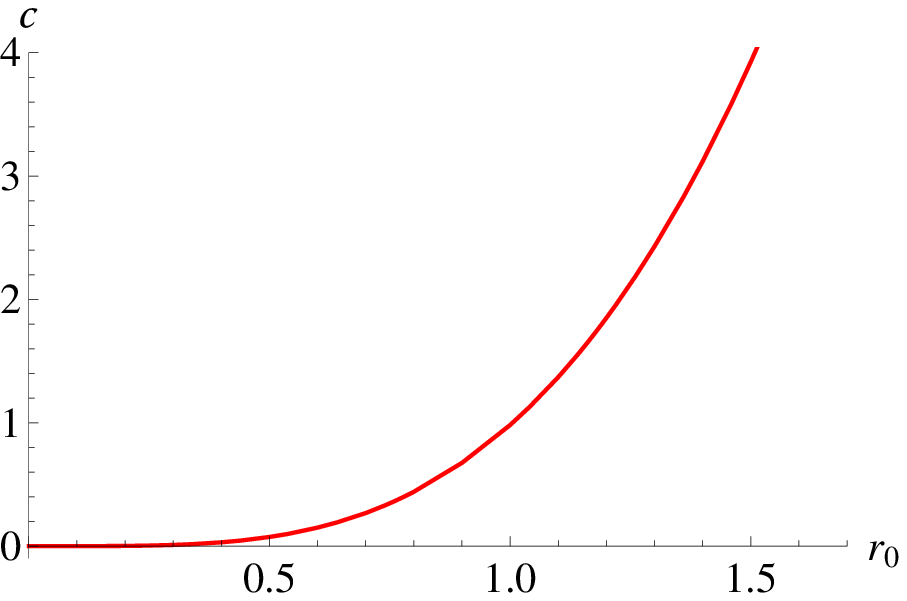}
\caption{{\small  $\Gamma$ for $E=2.0$, $m_q=1.0$ and $R=1$ are shown as the function of $r_0$ (left)
and of $c$ (middle). The right figure shows the relation of $c$ and $r_0$. 
 \label{prod-rate-ns}}}
\end{center}
\end{figure}

\subsection{\bf Field condensate and $\Gamma$}

As in the finite temperature case,
the production rate $\Gamma$ depends on the chiral condensate $c$, which is given
by the above profile function $w_0(\rho)$ as
\beq\label{wc}
 w_0(\rho)=m_q+{c\over \rho^2}+\cdots\, ,
\eeq
where $m_q$ denotes the current quark mass.

The estimated value of $\Gamma$ as a function of $c$
is shown in the Fig.\ref{prod-rate-ns} for $E=2.0$ and $m_q=1.0$, where
$\Gamma$ is also shown as a function of the parameter $r_0$.
$\Gamma$ decreases rapidly with $c$. This is reasonable since 
the effective mass $m_q^{eff}$ increases
with $c$ as shown below, and then $\Gamma$ decreases . 
As for $r_0$ dependence, $\Gamma$ is a decreasing function of $r_0$. This point is
understood since $c$ increases with $r_0$ as shown in the Fig. \ref{prod-rate-ns}.
We notice that $r_0$ is related to the condensate of gauge field strength 
$\langle F_{\mu\nu}F^{\mu\nu}\rangle\propto r_0^4$, which constructs the string tension
of the quark and anti-quark bound state.

\subsection{$m_q^{eff}$ and NJL coupling}

As for the estimation of $m_q^{eff}$, we can perform it
by changing the relation (\ref{meff}), which is given for the finite temperature case, as follows
\beq\label{meff-2}
    \Gamma(m_q, c)\equiv \Gamma_c(m_q)=\Gamma_0 (m_q^{eff})\, ,
\eeq
where the right hand side $\Gamma_0 (m_q^{eff})$ 
is considered to be the same with the one used in (\ref{meff})
since the limit of $T=0$ in the previous section and the limit of $r_0=0$ represent the same AdS$_5\times S^5$
bulk metric.

\begin{figure}[htbp]
\vspace{.3cm}
\begin{center}
 \includegraphics[height=5cm,width=5cm]{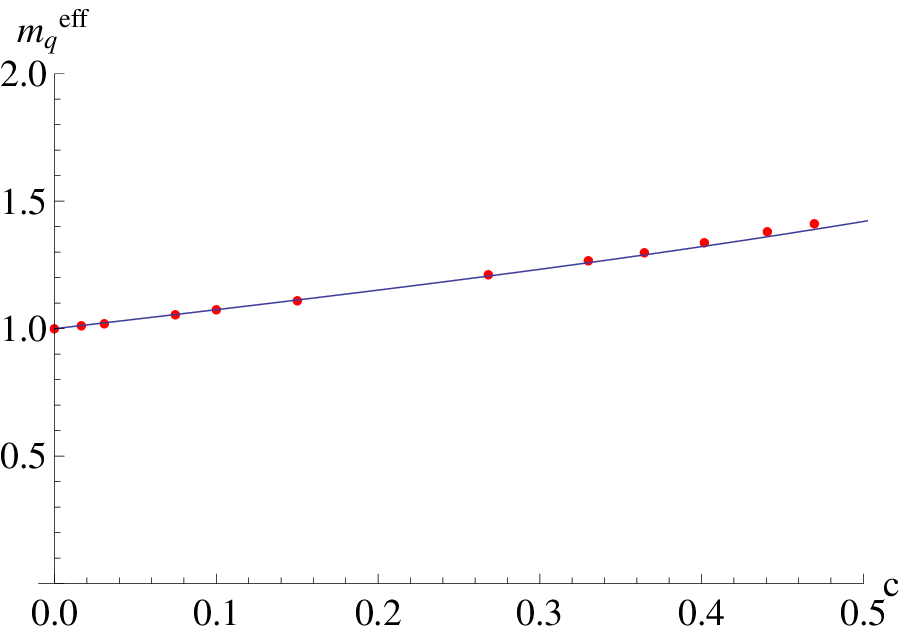}
\includegraphics[height=5cm,width=5cm]{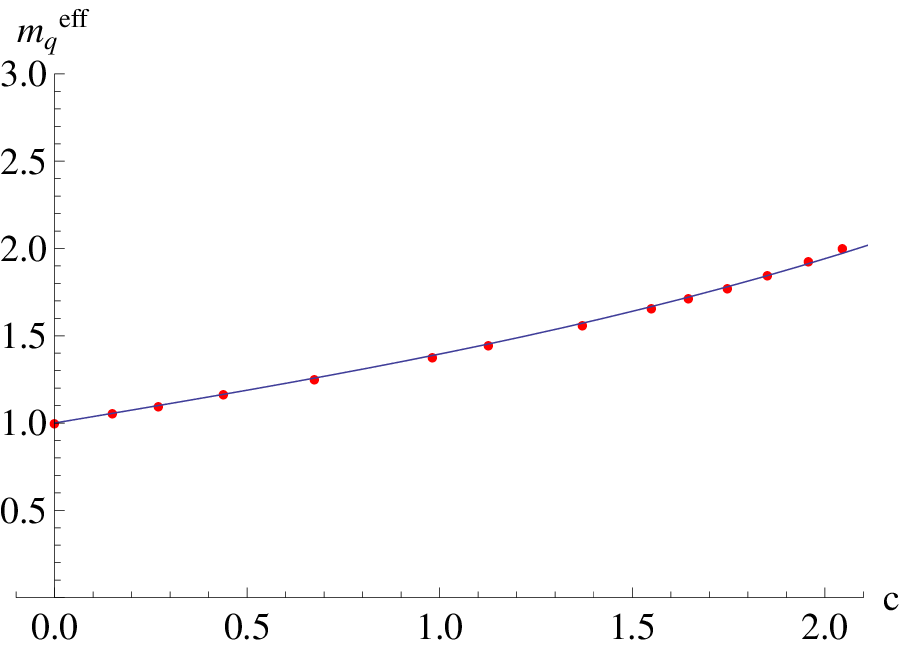}
\includegraphics[height=5cm,width=5cm]{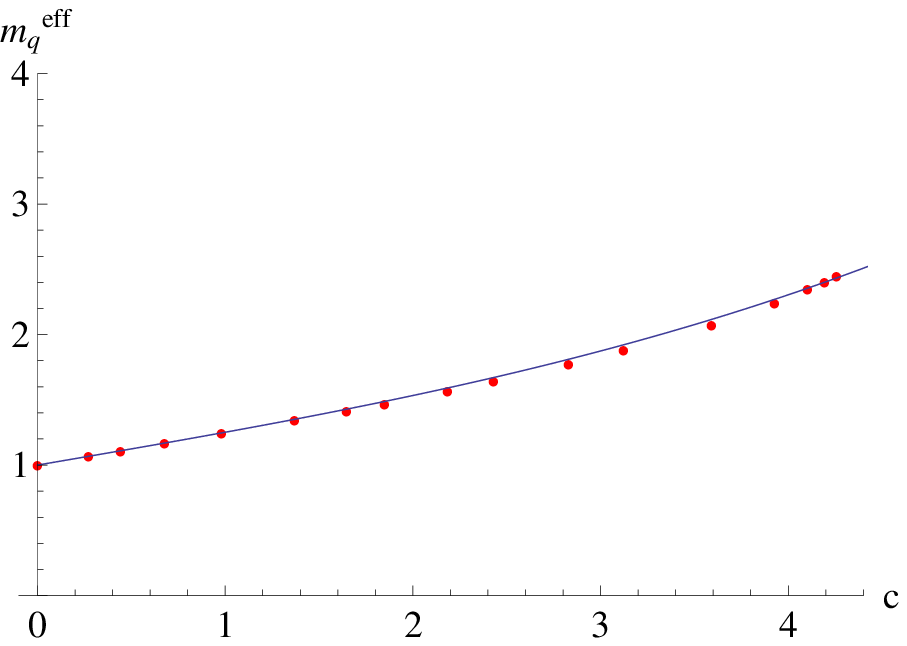}
 \caption{ Dots represnt the $m_q^{eff}$ for $R=1$ and $m_q=1$ with  $E=2$, $E=4$ and $E=6$ from the right respectively. The solid line
represents $m_q^{eff}=m_q+2g_s c+3h_2c^2+4h_3c^3$ with  
 $g_s=0.74/E$, $h_2=0$ and  $h_3=1.61/E^4$. \label{gs-c-1}}
\end{center}
\end{figure}

By using the above relation (\ref{meff-2}), the $m_q^{eff}$ is obtained as a function of $c$.
The results are shown in the Fig. \ref{gs-c-1} . Here  $m_q^{eff}$ runs from the current quark mass $m_q$ to the upper bound $R\sqrt{E}$ as  (\ref{bound}).

We compare our numerical results for $m_q^{eff}$ with the effective quark  mass from NJL model with higher order of $c$  
as follows, 
\beq\label{NJL-2}
  m_q^{eff}=m_q+2g_s c+3h_2c^2+4h_3c^3\,.
\eeq
By fitting the parameters numerically,  we obtain the coefficients scaled by the appropriate powers of $E$ as 
\beq \label{NJL-3}
  g_s=0.74/E\,, \quad h_2=0\,, \quad h_3=1.61/E^4\, .
\eeq
Here $m_q^{eff}$  can be fitted by including up to the term $c^3$. 
However, in this case, $h_2$ is not 
obtained in the expected form $const./E^{5/2}$, which is
scaled by the canonical power of $E$  stated in the section 3 through (\ref{scale-1}) and (\ref{scale-2}). Thus we get the reasonable solution such that $h_{2}=0$ as (\ref{NJL-3}). 
 This result seems to be inconsistent with the NJL model proposed
with higher order multi quark interactions in \cite{OHMBP}. 
We should notice that $c^2$ term in \cite{OHMBP} comes from six fermion interactions, which is introduced 
as the $U(1)_A$ breaking term with three flavor NJL model \cite{Kobayashi}. 
However, in our model, the flavor number $N_f$ is set as one. 
Thus, our result is consistent with this fact.



\section{Summary and Discussion}

In the D3/D7 brane system, a holographic Schwinger effect is studied  
by imposing an external electric field $E$ for ${\cal N}=2$ supersymmetric theory and also for two 
non-supersymmetric one, deconfining finite temperature and confining chiral symmetry breaking theory.  In the present approach, the quark pair production rate
 $\Gamma$ is given as the imaginary part of the on-shell D7-brane action. 

\vspace{.3cm}
For the deconfining and chiral symmetric
theory at finite temperature, the dual bulk background is given by $AdS_5$-Schwarzschild$\times S^5$  space-time. 
At zero temperature limit, the theory is reduced to the ${\cal N}=2$ supersymmetric theory, and we obtain the analytic form of production rate, $\Gamma_0(m_q)$. 
 The production rate obtained in this way is different from the one found via tunneling process
as Schwinger effect in QED. Our result gives a production rate via a vacuum decay process.

From a dynamical viewpoint, this point can be understood more precisely. In the present holographic theory,
the lower bound $E_c$ corresponds to
the tension of the linear potential, which is observed near very small distance 
between the quark and anti-quark for $E=0$. This is found in terms of 
the string whose endpoints on the D7-brane  are at the minimum point of $r$. 
This potential could make the bound state of the quark and anti-quark as mesons.
 The role of the imposed $E$ is to reduce this attractive force.
Actually, the condition, $E\ge E_c$, is sufficient to remove the attractive potential. 
 So any tunneling phenomenon can not be expected in this case 
for getting the pair production rate of quarks. 
As a result, 
the quark and anti-quark will be separated without any resistance under the imposed
$E$.

\vspace{.3cm}
For the finite temperature case, there exist two types of D7-brane embeddings,
the Minkowski and the BH types. For a fixed $m_q$, they are realized
at the low and high temperature respectively.   
The transition of the embedding type occurs at the temperature $T_c$, where   
we can see a gap of 
$\Gamma$ as shown in Fig.\ref{gm}. As for the value of $E_c$,
it is finite for the Minkowski type. On the other hand, it vanishes 
in the case of BH type. 
 This is understood as the reflection of the screening of the attractive force
due to the thermal fluctuation. 
Actually the meson states in the BH embeddings
are observed as the quasi-normal modes \cite{Sta}-\cite{eva}
which have complex frequency. Therefore they are unstable. This is the reason of $E_c=0$
in this case.

\vspace{.1cm}
Thus 
it is easier to create the quark pair in the phase of black hole type than in the case of the Minkowski type. This 
point is assured by the temperature dependence of $\Gamma$, which increases rapidly near the 
transition temperature from the Minkowski to the BH embedding. Above the critical temperature,
$\Gamma$ slowly approaches to the $\Gamma_0(0)$ 
with increasing temperature. 
In other words, the explicit breaking of the chiral symmetry due to the mass term is also restored at $T=\infty$ from the viewpoint of the effective quark mass $m_q^{eff}$
since $m_q^{eff}$ for all $m_q$ approaches to zero.

\vspace{.3cm}
The characteristic feature of the theory is found in 
$m_q^{eff}$. 
 Here $m_q^{eff}$ is pulled out 
by identifying  $\Gamma_T(m_q)$  with $\Gamma_0(m_q^{eff})$. 
The latter is the one of the supersymmetric theory dual to the AdS$_5\times S^5$ bulk and its analytic form is given.
Through this procedure,
we could see how $m_q^{eff}$ decreases with $T$. 
This behavior of $m_q^{eff}$ is also related to the chiral condensate $c$, which is negative finite
and decreasing with $T$. 
 For the Minkowski embedding, heavy meson states are still living in spite of the fact that the theory
is in the deconfinement phase. So,
we have tried to understand the behavior at finite $T$ from the NJL model by extending a simple 
mean field approximation. 
However the problem is not so simple. 
This problem  is left as a challenging task to build an effective NJL type model at finite temperature deconfining theory. 

 
\vspace{.3cm}
In the next, we extended the analysis to a quark confining and chiral symmetry breaking phase. As 
in the supersymmetric case, the lower bound
$E_c$ is determined by a linear 
 part of the potential at short distance, which is observed near very short distance between quark and anti-quark.
The important point is that the other linear potential is seen also in the long distance region in the present case.
This point is different from the case of the finite temperature deconfining phase. 
However, the $E_c$ is not altered since
the tension of the short range force is larger than the one of the long range confining force as shown by (\ref{tauqcd}).

The confinement of the present theory is supported
by the gauge field condensate $\langle F_{\mu\nu}F^{\mu\nu}\rangle\propto r_0^4$ which is characterized
by the parameter $r_0$. In this theory, the chiral symmetry is also broken. The order parameter $c$ is therefore positive
finite and increases with $r_0$. 

We find that 
the production rate $\Gamma$ is decreasing with $r_0$ as expected 
since the effective quark mass $m_{q}^{eff}$ increases with $c$.
In this case, the effective mass $m_{q}^{eff}$ is obtained by comparing $\Gamma$ at finite $r_0$ and the one at
$r_0=0$. We notice $\Gamma$ for $r_0=0$ is reduced to the one given at zero temperature AdS$_5\times S^5$ background. 
From the viewpoint of chiral symmetry breaking phase,
we studied the relation between the effective quark mass $m_{q}^{eff}$ and chiral condensate $c$. 
In this case, interestingly, we could get a consistent result with a simple NJL model with 4-th and 8-th terms of fermions.\\

Finally we comment on 
the tunneling process 
in 
the present D3/D7 case. 
We could find 
the tunneling via instanton configurations for $0<E<E_c$ in the deconfining phase for a fixed $m_q$. 
In the confining phase, we would need a new lower bound  $E_s(=\tilde{\tau}_{QCD})$ to overcome the confining force at long distance
in order to realize the tunneling pair production. Then the tunneling process is found for $E_s<E<E_c$.
 In the case of $E>E_c$,  on the other hand, we will find the tunneling production rate by using
the real part of the D7 brane action. we will discuss these points in the future paper.


\newpage

\vspace{.3cm}
\section*{Acknowledgments}
The authors would like to thank K. Hashimoto for useful discussions. M. Ishihara would like to thank H. Suganuma and K. Kashiwa  for useful discussions.
The work of M. I. is supported
by World Premier International Research Center Initiative WPI, MEXT, Japan. The work of M. I. is supported in part
by the JSPS Grant-in-Aid for Scientific Research, Grant No. 15K20877.



\newpage
\end{document}